\documentclass[10pt,twocolumn,letterpaper]{article}

\usepackage[pagenumbers]{cvpr} 

\usepackage[dvipsnames]{xcolor}

\definecolor{cvprblue}{rgb}{0.21,0.49,0.74}
\usepackage[pagebackref,breaklinks,colorlinks,citecolor=cvprblue]{hyperref}

\usepackage{graphicx}
\usepackage{multirow}
\usepackage{multicol}
\usepackage{bm}
\usepackage{balance}
\usepackage{bbding}

\def\paperID{14157} 
\def\confName{CVPR}
\def\confYear{2024}

\title{Auffusion: Leveraging the Power of Diffusion and Large Language Models for Text-to-Audio Generation}
\author{
Jinlong Xue, Yayue Deng, Yingming Gao, Ya Li\thanks{Corresponding author.} \\
Beijing University of Posts and Telecommunications\\
{\tt\small \{jinlong\_xue, yayue.deng, yingming.gao, yli01\}@bupt.edu.cn}
}

\begin{document}
\maketitle
\vspace{-0.2cm}
\begin{abstract}

Recent advancements in diffusion models and large language models (LLMs) have significantly propelled the field of AIGC. Text-to-Audio (TTA), a burgeoning AIGC application designed to generate audio from natural language prompts, is attracting increasing attention. However, existing TTA studies often struggle with generation quality and text-audio alignment, especially for complex textual inputs. Drawing inspiration from state-of-the-art Text-to-Image (T2I) diffusion models, we introduce Auffusion, a TTA system adapting T2I model frameworks to TTA task, by effectively leveraging their inherent generative strengths and precise cross-modal alignment. Our objective and subjective evaluations demonstrate that Auffusion surpasses previous TTA approaches using limited data and computational resource. Furthermore, previous studies in T2I recognizes the significant impact of encoder choice on cross-modal alignment, like fine-grained details and object bindings, while similar evaluation is lacking in prior TTA works. Through comprehensive ablation studies and innovative cross-attention map visualizations, we provide insightful assessments of text-audio alignment in TTA. Our findings reveal Auffusion’s superior capability in generating audios that accurately match textual descriptions, which further demonstrated in several related tasks, such as audio style transfer, inpainting and other manipulations. Project page is available at \href{https://auffusion.github.io}{https://auffusion.github.io}.

\vspace{-0.3cm}

\end{abstract}    
\section{Introduction}
\label{sec:intro}

Text-to-audio (TTA) generation is an emerging application that focuses on synthesizing diverse audio outputs based on text prompts. With the integration of artificial intelligence into the realm of AIGC, the scope of TTA applications has expanded significantly, covering areas such as movie dubbing and musical composition. Early TTA models, as referenced in \cite{DBLP:conf/mlsp/LiuIZHPW21,DBLP:conf/icassp/KongXICWP19}, primarily relied on one-hot labels, leading to the generation of monotonous audio constrained by limited label space and generative capacity. In contrast, natural descriptive text delivers more comprehensive and fine-grained information. Thereby, the following works~\cite{Diffsound,AudioLDM,AudioLDM-2,tango,Make-An-Audio,make-an-audio2} develop their models based on textual content. 

Recent advancements in diffusion generative models \cite{DBLP:conf/icml/Sohl-DicksteinW15,DDPM} and large language models \cite{t5,roberta,clip,clap,flant5} have showcased remarkable capabilities in content generation and understanding. Leveraging these advancements, the first diffusion based TTA Diffsound \cite{Diffsound} outperforms previous TTA systems by generating discrete tokens quantized from mel-spectrograms using a diffusion model. Later Diffsound is surpassed by AudioGen~\cite{audiogen} using an autoregressive model in a discrete space of waveform. Inspired by \cite{stablediffusion}, AudioLDM~\cite{AudioLDM} is the first to utilize a continuous latent diffusion model (LDM) \cite{stablediffusion}, achieving better quality and computational efficiency compared to other discrete token based TTA systems \cite{Diffsound,audiogen}. Similarly, many well-performed TTA systems, including AudioLDM2~\cite{AudioLDM-2}, Tango \cite{tango}, Make-an-Audio 1/2 \cite{Make-An-Audio,make-an-audio2}, integrate LDM into their TTA frameworks to facilitate denoising processes in the latent space. However, these models still require extensive computational resources and large-scale datasets for training from scratch. Moreover, these models only emphasis coarse-grained performance and neglect fine-grained text-audio alignment. Our work concentrates on addressing these two critical challenges, providing effective solutions and valuable insights.

In developing a powerful TTA model, two primary objectives are paramount: 1) mastering the distribution of natural audio, and 2) achieving precise text-audio alignment. These competencies are also paralleled in T2I~\cite{imagen, PIXART} tasks, where robust generative abilities and accurate text-image alignment are similarly crucial. To this end, we introduce our TTA system Auffusion that adapts Latent Diffusion Model (LDM) originally pretrained for T2I tasks. This adaptation enables Auffusion to leverage the LDM's inherent generative strength and transfer it alignment understanding effectively for text-audio alignment in TTA application. The comprehensive subjective and objective evaluation metrics show our proposed system Auffusion achieves better quality and alignment. Moreover, Auffusion demonstrates performance comparable to other baseline models trained on datasets 60 times larger. Riffusion \cite{riffusion}, an app for 5-second music generation, also uses a pretrained LDM \cite{stablediffusion}. However, it is specifically designed for music and does not extend to broader TTA tasks. Additionally, Riffusion simply quantizes audio signals into images, a non-reversible process that leads to a great precision loss. In contrast, our Auffusion model features a carefully designed feature space transformation pipeline, enabling lossless audio conversion.

On the other hand, the text encoder serves as a critical bridge between text and audio, representing a key component in TTA systems. Different from the extensive studies~\cite{imagen,ediff,t2i-compbench} conducted in the T2I domain, where the impact of text encoders on aspects such as fine-grained details and object bindings has been widely explored, their influence in TTA has not been thoroughly explored. Generally, text encoders in existing TTA systems fall into two categories: 1) multi-modal contrastive learning-based models, such as CLIP \cite{clip} and CLAP \cite{clap}, and 2) text-only large language models (LLMs) like BERT \cite{bert} and T5 \cite{t5}. However findings from previous studies can sometimes be contradictory. Diffsound \cite{Diffsound} employs CLIP, pretrained on text-image pairs, claiming superior performance over text-only model BERT. Conversely, AudioLDM \cite{AudioLDM} uses CLAP model, pretrained on text-audio pairs, suggesting advantages using audio only features over using combined audio-text features or text-only features. Building upon the same LDM used in AudioLDM, Tango \cite{tango} owes different opinions. They advocate for instruction-tuned LLMs (Flan-T5)~\cite{flant5} to better grasp textual descriptions and cross-modal correlations, challenging the notion of embedding audio and text in a shared space. However, comprehensive ablation studies using these text encoders are lacking.

To address the debates outlined above, our study conducts a thorough investigation into the performance of various text encoders and baseline models. Moving beyond traditional evaluation metrics, we innovatively assess text-audio alignment by visualizing the cross-attention map to provide intuitive observation for the first time in TTA task. We find that our model achieves better fine-grained text-audio alignment. Additionally, we demonstrate versatile audio manipulations enabled by our model's generative capacity and clear text-audio alignment.

In summary, the contributions of our work are: 1) We propose Auffusion, a TTA model that integrates a powerful pretrained LDM from T2I in order to inherit generative strengths and enhance cross-modal alignment, and our method demonstrates superior performance compared to existing TTA systems; 2) We conduct an extensive investigation into the performance between different text encoders, and we provide a novel and insightful demonstration in TTA task to assess text-audio alignment utilizing visualizations of cross-attention maps across different models.
\section{Related Work}
\label{sec:formatting}

\subsection{Text-to-Image Synthesis}
Text-to-Image (T2I) synthesis, particularly through diffusion models, has seen significant advancements in recent years. Pioneering models like DALL-E~\cite{dalle} treats T2I as a sequence-to-sequence translation task, encoding images into discrete latent tokens using pretrained VQ-VAE~\cite{vq-vae}. DALL-E 2~\cite{dalle2} employs the CLIP text encoder and two diffusion models, first predicting CLIP~\cite{clip} visual features from text and then synthesizing the image. Another famous model Imagen~\cite{imagen} uses the T5 encoder~\cite{t5} for text feature extraction and a cascade of diffusion models for initial image synthesis and subsequent super-resolution. Stable Diffusion~\cite{stablediffusion} optimizes computational efficiency by mapping images from pixel to compressed latent space using a continuous VAE trained with discriminators, followed by diffusion in this latent space. These models~\cite{glide,stablediffusion,imagen,dalle2} demonstrate remarkable diversity and quality in image generation, guided by text prompts and operating either directly in image space or within a latent space.

\subsection{Text-to-Audio Synthesis} 
Text-to-Audio (TTA) synthesis has witnessed significant advancements. Diffsound~\cite{Diffsound} leverages VQ-VAE~\cite{vq-vae} model trained on mel-spectrograms and convert them into discrete codes, where a non-autoregressive token based diffusion model is then used to generate audio signals. Similarly, AudioGen~\cite{audiogen} uses a VQ-VAE based approach~\cite{encodec} but focuses on encoding raw waveform data and employs an autoregressive model for generation. Other studies include the use of Latent Diffusion Models (LDMs), as seen in AudioLDM 1/2~\cite{AudioLDM,AudioLDM-2}, Make-An-Audio 1/2~\cite{Make-An-Audio,make-an-audio2}, and Tango~\cite{tango}. AudioLDM~\cite{AudioLDM}, utilizes audio features extracted by a pretrained contrastive text-audio model CLAP~\cite{clap} as a condition during training, while leveraging text features during inference. This approach benefits from CLAP's ability to map audio and captions to a shared latent space. AudioLDM2~\cite{AudioLDM-2} first employs an auto-regressive model (AR) to generate AudioMAE~\cite{audiomae} features from text, then uses them to condition the LDM. These two methods both alleviate the reliance of audio-text pair data. Other methods~\cite{Make-An-Audio,make-an-audio2,tango}, on the other hand, employ text features in both training and inference stages. Make-An-Audio~\cite{Make-An-Audio}' LDM is similar to AudioLDM. Make-An-Audio2~\cite{make-an-audio2} emphasize the temporal information by changing 2D spatial structure to 1D temporal structure, and they additionally replace U-Net design to transformer. However, neither of these two models is open source.

\begin{figure*}[ht!]
  \centering
   \includegraphics[width=\linewidth]{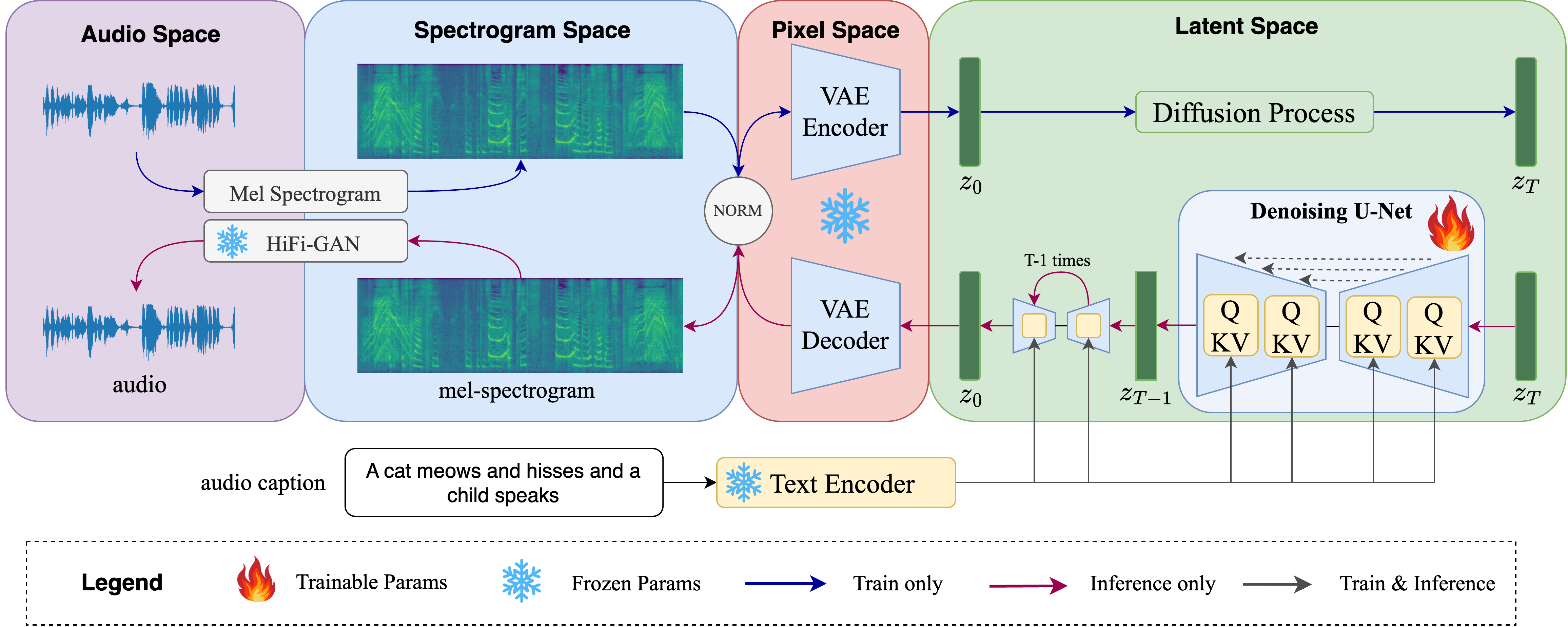}
   \caption{An overview of Auffusion architecture. The whole training and inference process include back-and-forth transformation between four feature spaces: audio, spectrogram, pixel and latent space. Note that U-Net is initialized with pretrained text-to-image LDM.
   }
   \label{fig:overview}
\end{figure*}

\section{Auffusion}

\subsection{Overview}

Our proposed method Auffusion, as depicted in Fig.~\ref{fig:overview} has four main components: 1) text encoder; 2) latent diffusion model (LDM); 3) pixel VAE; 4) HiFi-GAN vocoder. In order to achieve the TTA task and utilize the powerful pretrained model from T2I task, the whole process involves conversion between four feature spaces: audio, spectrogram, pixel and latent space. The spectrogram feature is a key proxy that bridges the audio space and pixel space. During training, audio is first converted into mel-spectrogram and normalize for image space, then LDM is conditioned on textual embeddings extracted by textual condition encoder and trained in the pixel space learned by VAE. In inference, this process is reversed: starting from standard Gaussian noise, the latent representation is generated through reverse diffusion process conditioned on text embeddings. Thereafter the pixel VAE decoder reconstructs the pixel space and generated image is denormalized into mel-spectrogram. Finally, the pretrained HiFi-GAN vocoder synthesizes the audio from this mel-spectrogram.

\subsection{Feature Space Transformation}
Given an audio-text pair, we first convert the audio signal $x_{audio} \in \mathbb{R}^{T}$ into mel-spectrogram $x_{mel} \in \mathbb{R}^{d\times l}$, where $d$ and $l$ represent the mel-channels and the number of frames respectively. In order to transform mel-spectrogram into image-like data without precision loss, we conduct normalization by utilizing the mean $\mu$ and variance $\delta$ calculated from the whole dataset rather than on individual mel-spectrogram~\cite{riffusion}. The normalized spectrogram $x_{norm}$ can be viewed as gray-scale image and then converted into RGB image data $x \in \mathbb{R}^{c \times d \times l}$, where $c$, $d$ and $l$ are referred to as the image channel, height and width respectively.

\subsection{Latent Diffusion Model}

To guide the construction of the audio signal's pixel distribution $z_0$ using a text prompt $\tau$, we fine-tune the U-Net diffusion module by minimizing mean squared error (MSE) in the noise space. The objective function is defined as:
\begin{equation}
\ell_\theta = ||\epsilon_\theta(z_t, t, \tau) - \epsilon||^2_2
\end{equation}
Here, $\epsilon \sim \mathcal{N}(0, I)$ represents Gaussian noise, $t$ is a random time step, and $\epsilon_\theta$ is the text-guided denoising network, comprising a U-Net with a cross-attention component for text guidance $\tau$.

In this process, the VAE encoder processes the image-like input $x$ into compressed latent vector $z_0$. The diffusion process then operates in this latent space, gradually transforming $z_0$ into Gaussian noise $z_T$. The model is trained to reverse this transformation, recovering the original data distribution from the noise. This process involves two key steps: the forward process that converts $z_0$ into $z_T$ and the reverse process that recovers $z_0$ from $z_T$.

\textbf{Forward process} is defined by a fixed Markov chain from data {\tt $z_0$} to the latent variable {\tt $z_T$}.
\begin{equation}
  q(z_1,\dots,z_T|z_0) := \prod_{t=1}^T q(z_t|z_{t-1}) 
  \label{eq:forward}
\end{equation}
The entire procedure transforms the initial latent data {\tt $z_0$} into noise latent variables {\tt $z_T$} in accordance with a predetermined noise schedule $\beta_1,\dots,\beta_T$.
\begin{equation}
    q(z_t|z_{t-1})  := \mathcal{N}(z_t;\sqrt{1-\beta_t} z_{t-1},\beta_t I)
\end{equation}
where $\beta_t$ is a small positive constant. $q(z_t|z_{t-1})$ represents a function where a small Gaussian noise is added to the distribution of $z_{t-1}$.

\textbf{Reverse process} converts the latent variables from $z_T$ to $z_0$ with learnable parameters $\theta$, aimed at recovering samples from Gaussian noise $z_T\sim \mathcal{N}(0,I)$. 
\begin{align}
    p_\theta(z_0,\dots,z_{T-1}|z_T) := \prod_{t=1}^Tp_\theta(z_{t-1}|z_t) \\
    p_\theta(z_{t-1}|z_t):=\mathcal{N}(z_{t-1};\mu_\theta(z_t,t),\Sigma_\theta(z_t, t))    
\end{align}

Note that $\mu_\theta$ takes the diffusion step $t \in \mathbb{N}$ and variable $z_t$ as inputs and outputs $z_{t-1}$ for each iteration.

\vspace{0.3cm}

\subsection{Conditioning Processes}
\label{sec:conditioning}
The previous work AudioLDM \cite{AudioLDM} adopts concatenation operation between pooled text embedding extracted from CLAP and time embedding to guide the generation process in LDM. By contrast, we turn diffusion model generation into more flexible and understandable by conducting a cross-attention mechanism \cite{attention} between conditional embedding sequence and latent vectors in the U-Net backbone. 

More formally, we donates $\vartheta_i(z_t) \in \mathbb{R}^{N \times d^i_\epsilon}$ a intermediate representation of the $i$\textit{-th} layer of U-Net estimation $\epsilon_\theta$. Then, a linear projection is applied to the deep spatial features of the noisy data $\vartheta_i(z_t)$.
\begin{equation}
    Q = W_q^{(i)} \cdot \vartheta_i(z_t)
\end{equation}
The conditional embedding $\tau$ is also projected via learned linear projections.
\begin{equation}
    K = W_k^{(i)} \cdot \tau,\quad
    V = W_v^{(i)} \cdot \tau,    
\end{equation}
where $W_q^{(i)} \in \mathbb{R}^{d\times d_\tau}$, $W_v^{(i)} \in \mathbb{R}^{d\times d_\epsilon^i}$ and  $W_k^{(i)} \in \mathbb{R}^{d\times d_\tau}$ are learnable matrices. The attention value and attention score are calculated as follows:
\begin{equation}
\begin{split}
    &Attention(Q,K,V) = score(Q,K) \cdot V \\
    &score(Q,K) = softmax (\frac{QK^T}{\sqrt{d}})
\end{split}
\end{equation}
The condition approach allows us to visualize the 2D attention map~\cite{prompt2prompt,inversion,daam} by reshaping attention score back to latent image shape. Furthermore, it provides an intuitive measurement to access the understanding ability of various text encoders. We discuss the compared results in Sec.~\ref{sec:analysis}. Meanwhile, based on the visualized attention map, we find that the pretrained LDM is capable of adequately transferring cross-modal understanding ability from T2I to TTA task, resulting in better alignment. Overall, we highlight the importance of the conditioning process in enhancing the audio-text model's ability to extract key information from text descriptions and accurately match the desired audio, as demonstrated in Fig.~\ref{fig:attn_map}.

\subsection{Text Encoder}
Inspired by eDiff-I~\cite{ediff} who uses an ensemble of encoders to provide multi-source information to LDM, we combine CLAP and FlanT5 text encoders as conditions. We use random dropout on each of these embeddings independently during training. When all two embeddings are dropped, it corresponds to unconditional training, which is useful for performing classifier-free guidance~\cite{cfg}. We conduct comprehensive comparison for various text encoders and results are shown in Sec~\ref{sec:analysis}. 

\subsection{Classifier-Free Guidance}
To guide the reverse diffusion process, we utilize classifier-free guidance~\cite{cfg} based on the text input $\tau$ using: 
\begin{equation}
    \hat{\epsilon}_\theta(z_t,t,\tau) = (1+w) \cdot \epsilon(z_t,t,\tau) - w \cdot \epsilon(z_t,t)
\end{equation}
At the inference stage, the guidance scale $w$ determines how much the text input influences the noise estimation $\hat{\epsilon}_\theta$ compared to the unconditional estimation. We randomly discard the text condition at a rate of 10\% during training.

\section{Experiments}

\subsection{Experimental Setup}

\begin{table*}[ht!]
\centering
\caption{The comparison between our model Auffusion and baseline TTA models. Although our model Auffusion is only trained on a much smaller dataset AC, our model outperforms other baselines on AC test set and has comparable zero-shot result in Clotho test set.}
\label{tab:overall_result}
\scalebox{0.82}{
\begin{tabular}{l|c|c|c|ccccc|ccccc}
\toprule
\multirow{2}{*}{Model} & \multirow{2}{*}{Pretrain} & \multirow{2}{*}{Duration(h)} & \multirow{2}{*}{Params} & \multicolumn{5}{c|}{\textbf{AudioCaps}} & \multicolumn{5}{|c}{\textbf{Clotho}} \\
 &  & &  & {FD$\downarrow$} & {FAD$\downarrow$} & {KL$\downarrow$} & {IS$\uparrow$} & {CLAP$\uparrow$} & {FD$\downarrow$} & {FAD$\downarrow$} & {KL$\downarrow$} & {IS$\uparrow$} & {CLAP$\uparrow$} \\
\midrule
Riffusion          &\CheckmarkBold & 1990 & 1.1B & 26.28 & 4.68 & 1.57 & 7.21 & 47.6\% & 31.63 & 6.11 & 2.66 & 6.50 & 47.5\% \\
AudioGen-v2-medium &\XSolidBrush & 6824 & 1.5B & \textbf{17.86} & 1.73 & 1.59 & 9.31 & 48.5\% & 23.26 & 2.55 & 2.56 & 7.19 & 46.7\% \\
AudioLDM-S-full-v2 &\XSolidBrush & 9031 & 421M & 30.58 & 4.40 & 1.79 & 6.96 & 42.1\% & 26.51 & 3.54 & 2.62 & 6.58 & 49.9\% \\
AudioLDM-L-full    &\XSolidBrush & 9031 & 975M & 29.77 & 4.04 & 1.78 & 7.50 & 42.9\% & 24.13 & 3.02 & 2.56 & 7.49 & 51.0\% \\
AudioLDM2          &\XSolidBrush & 29510 & 1.1B & 26.05 & 1.94 & 1.76 & 7.31 & 46.0\% & 23.53 & 3.06 & 2.47 & 9.05 & 48.1\% \\
AudioLDM2-large    &\XSolidBrush & 29510 & 1.5B & 25.59 & 2.19 & 1.70 & 7.83 & 47.9\% & 23.31 & 3.00 & 2.41 & 8.88 & 49.0\% \\
Tango              &\XSolidBrush & 145 & 1.3B & 24.82 & 1.77 & 1.43 & 7.20 & 55.0\% & 31.67 & 3.22 & 2.57 & 7.18 & 46.6\% \\
Tango-Full         &\XSolidBrush & 3400 & 1.3B & 30.68 & 3.67 & 1.63 & 4.79 & 51.9\% & 25.83 & 3.17 & \textbf{2.35} & 6.51 & 50.3\% \\
\midrule
Auffusion       &\XSolidBrush & 145 & 1.1B & 24.45 & 2.25 & 1.39 & 10.14 & 54.7\% & 29.01 & 2.67 & 2.66 & 9.46 & 47.6\% \\
Auffusion       &\CheckmarkBold & 145 & 1.1B & 21.99 & \textbf{1.63} & \textbf{1.36} & \textbf{10.57} & 55.3\% & 25.64 & 2.35 & 2.59 & 9.01 & 48.2\% \\
\midrule
Auffusion-Full  &\XSolidBrush & 1990 & 1.1B & 24.11 & 1.67 & 1.46 & 8.39 & 51.6\% & 19.14 & 1.99 & 2.42 & 10.33 & 52.8\% \\
Auffusion-Full  &\CheckmarkBold & 1990 & 1.1B & 23.08 & 1.76 & \textbf{1.36} & 10.28 & \textbf{55.6\%} & \textbf{17.97} & \textbf{1.96} & 2.38 & \textbf{11.29} & \textbf{55.0\%} \\

\bottomrule
\end{tabular}
}
\end{table*}

\textbf{Dataset.} We follow previous works~\cite{tango, make-an-audio2, AudioLDM} and use a variety of different audio datasets with audio caption or audio labels to train our model, including AudioCaps (AC)~\cite{audiocaps}, WavCaps~\cite{wavcaps}, MACS~\cite{macs}, Clotho~\cite{clotho}, ESC50~\cite{esc50}, UrbanSound~\cite{urbansound}, Music Instruments dataset\footnote{https://www.kaggle.com/datasets/soumendraprasad/musical-instruments-sound-dataset} and GTZAN~\cite{gtzan}. The WavCaps dataset consists of ChatGPT-assisted weakly-labeled audio captions for the FreeSound\footnote{https://freesound.org/}, BBC Sound Effects (SFX)\footnote{https://sound-effects.bbcrewind.co.uk/}, SoundBible\footnote{https://soundbible.com/} and the AudioSet strongly labeled subset~\cite{audioset-sl}, containing 403,050 audio clips with an average duration of 68 seconds. AudioCaps is a subset of AudioSet (AS)~\cite{audioset} with handcrafted captions and it contains about 46K ten-second audio clips. This results in a dataset composed of 0.47 million audio text pairs, with a total duration of approximately 7.7K hours.

It is noted that the duration of the audio samples in AudioSet and AudioCaps is 10 seconds, while it is much longer in FreeSound and BBC SFX datasets (86s and 115s in average). To avoid the imbalance caused by longer audio, which often contains repeated sounds like background sounds, we only use the first thirty seconds of audios for all datasets and randomly select ten-second segments during training. Finally, we have in total 0.4M audio samples with a total duration of around 2K hours for model training.

\textbf{Training Setup.} We utilize the pretrained Stable Diffusion v1.5\footnote{https://huggingface.co/runwayml/stable-diffusion-v1-5}, including its VAE and U-Net, and later finetune the U-Net on audio datasets. All datasets are resampled to 16kHz sampling rate and mono format, with samples padded to 10.24 seconds. We then extract mel-spectrograms from audios using parameters of 256 mel filter bands, 1024 window length, 2048 FFT, and 160 hop size, resulting in (1,256,1024) mel-spectrograms, akin to grayscale images with 256 height and 1024 width in 1 channels. These are normalized and channel-repeated to create RGB-like images suitable for VAE encoder input. For high-fidelity audio conversion, previous methods adopt neural vocoder~\cite{hifigan,bigvgan}. In order to match our need using our specific mel-spectrogram parameters, we train a new HiFi-GAN vocoder~\cite{hifigan} using the same datasets described above. This training employs the AdamW optimizer~\cite{adamw} with a 2e-4 learning rate and 16 batch size on one A6000 GPU. Finally, we freeze the text encoder and finetune the pretrained Stable Diffusion's U-Net using AdamW optimizer with a 3e-5 learning rate, at a 20 batch size for 100K steps. Our model can be trained only taking a total of 48 hours on one A6000 GPU.

\textbf{Objective Evaluation.} In our experimental evaluation, we follow previous evaluation methods~\cite{AudioLDM,tango,make-an-audio2} and employ a suite of objective metrics to assess the quality and fidelity of generated audio samples, including Frechet Distance (FD), Frechet Audio Distance (FAD), Kullback–Leibler (KL) divergence, Inception Score (IS) and CLAP score. Analogous to the Frechet Inception Distance (FID)~\cite{fid} used in image synthesis, the FD score in audio domain quantifies the global similarity between created audio samples and the target samples without the need of using paired reference audio samples. The IS score is effective at assessing both the quality and variety of samples. The KL score is calculated using paired samples and it measures the divergence between two probability distributions. These three metrics are all grounded in the advanced audio classifier PANNs~\cite{panns}. FAD score has a similar idea to FD but it uses VGGish~\cite{vggish} as feature extractor. The evaluate suite that we uses for FD, FAD, KL and IS is in project\footnote{https://github.com/haoheliu/audioldm\_eval}. Besides, we also use pretrained CLAP\footnote{https://huggingface.co/laion/clap-htsat-unfused} model to compute the similarity of the text caption and generated audio to evaluate the text-audio alignment, similar to CLIP score.

\textbf{Subjective Evaluation.} Following previous evaluation method~\cite{tango,audiogen} in TTA field, we ask five human evaluators to assess two aspects of the generated audio, including overall audio quality (OVL) and relevance to the text caption (REL). We randomly select 30 audio samples from each of the AC and Clotho test sets and ask participants to rate them on a scale from 1 to 100 with 10-point intervals. Results are shown in Table~\ref{tab:subjective_evaluation}.

\textbf{Baseline Models.} To comprehensively compare our models with others, our study employs five baseline models, including three diffusion based models Riffusion~\cite{riffusion}, AudioLDM~\cite{AudioLDM}, AudioLDM2~\cite{AudioLDM-2}, Tango~\cite{tango}, and one auto-regressive generative model based on discrete audio token AudioGen~\cite{audiogen}. We re-implement Riffusion\footnote{https://huggingface.co/riffusion/riffusion-model-v1}, originally trained on music datasets for only 5s audio, to generate 10s audio using a 160 hop length and trained it on our datasets. We use the other baseline models released by authors on huggingface respectively. AudioLDM-S-full-v2\footnote{https://huggingface.co/cvssp/audioldm-s-full-v2} and AudioLDM-L-full\footnote{https://huggingface.co/cvssp/audioldm-l-full} have 412M and 975M parameters and trained them on AudioCaps, AudioSet and other 2 datasets including 9031h audio data for more than 1.5M train steps. AudioLDM2\footnote{https://huggingface.co/cvssp/audioldm2} and AudioLDM2-large\footnote{https://huggingface.co/cvssp/audioldm2-large} have 1.1B and 1.5B parameters respectively and trained on 29510h diverse audio data. Tango is trained on AudioCaps dataset and Tango-Full is trained on datasets similar with our datasets settings but using different preprocessing. AudioGen-v2-medium\footnote{https://huggingface.co/facebook/audiogen-medium} has 1.5B parameters and is trained on AudioCaps, AudioSet, and eight other datasets around 4K hours data.

\begin{figure*}[ht!]
  \centering
   \includegraphics[width=\linewidth]{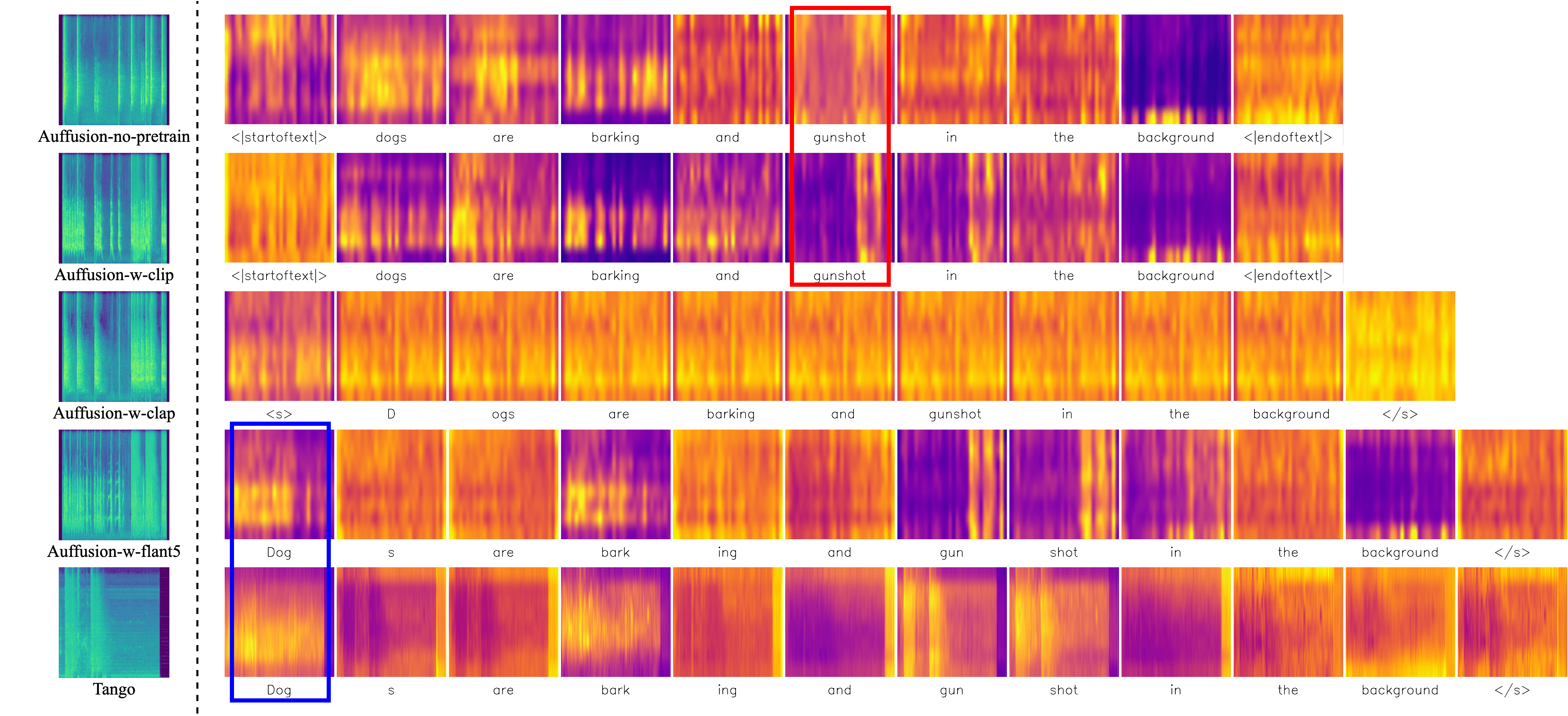}
   \caption{The visualization of cross attention maps for Auffusion with different text encoders and Tango model. Auffusion-no-pretrain use fixed CLIP encoder and LDM is trained from scratch. The LDMs in 2 to 4 rows are initialized with SDv1.5 with different encoders. The last row shows the Tango's cross attention map, and Tango uses FlanT5-large as condition encoder.}
   \label{fig:attn_map}
\end{figure*}

\subsection{Results}

\textbf{Evaluation Setup.} We compare our model Auffusion trained on single dataset AudioCaps (AC) and Auffusion-Full trained on whole datasets with other baselines in both AudioCaps test set and Clotho test set. We also conduct ablation studies on the impact of pretrained SD models. Both Auffusion and Auffusion-Full use CLIP as default text encoder. We report our result in Table~\ref{tab:overall_result}.

\textbf{Objective Evaluation.} When trained solely on AC dataset, our model Auffusion-with-pretrained outperforms the previous state-of-the-art Tango in AC test set, with 21.99 FD, 1.63 FAD, 1.36 KL, 10.57 IS and 55.3\% in CLAP score, and achieve comparable zero-shot Clotho test set results to other baseline models trained on datasets over 60 times larger. Notably, Tango and Auffusion-no-pretrained both trained solely on AC dataset exhibit a huge drop on Clotho test set, indicating a problem of overfitting. In contrast, our Auffusion-with-pretrained still maintain its performance, demonstrating generalization ability. This suggests that the generative capacity and cross-modal alignment of the pretrained SD model can be effectively transferred to mel-spectrogram domain, even with a small dataset.

When trained in much larger datasets using same training steps, Auffusion-Full-with-pretrained achieves the state-of-the-art performance in Clotho test set. Besides, it shows a negligible decrease in AC test set compared to Auffusion-with-pretrained, and records a slight increase in CLAP score. This indicates the robustness and strong generalization ability of the pretrained SD, even when dealing with different data distributions. Additionally, both our Auffusion-with-pretrained and Auffusion-Full-with-pretrained models significantly outperform other baselines in terms of IS and CLAP scores. A higher IS score implies that our model can generate mel-spectrograms with both high fidelity and diversity. A higher CLAP score indicates our model's enhanced capability to adhere to textual descriptions and produce more relevant audio. We also find that our re-implement Riffusion yields much inferior results, indicating that the precision loss caused by the quantization transform has a great impact.

\textbf{Subjective Evaluation.} Our subjective human evaluation results are presented in Table~\ref{tab:subjective_evaluation}. In the ``All Event" column, our model Auffusion-w-clip demonstrates superior performance over other baseline models, achieving an OVL score of 69.36 and a REL score of 70.25. Additionally, the REL has significant gains compared to other models, showing strong text-audio alignment. We delve deeper into the impact of numbers of events and provide intuitive visualization for text-audio alignment in Sec.~\ref{sec:analysis}.

\begin{table*}[ht!]
\centering
\caption{The results evaluated on AudioCaps test set and Clotho test set with different settings of conditional encoder.}
\label{tab:different_encoder}
\scalebox{0.85}{
\begin{tabular}{@{}l|ccccc|ccccc}
\toprule
& \multicolumn{5}{|c|}{\textbf{AudioCaps}} & \multicolumn{5}{|c}{\textbf{Clotho}} \\
\cmidrule{2-11}
\textbf{Auffusion-Full-with} & FD$\downarrow$ & FAD$\downarrow$ & KL$\downarrow$ & IS$\uparrow$ & CLAP$\uparrow$ & FD$\downarrow$ & FAD$\downarrow$ & KL$\downarrow$ & IS$\uparrow$ & CLAP$\uparrow$ \\
\midrule
CLIP & 23.08 & 1.76 & 1.36 & \textbf{10.28} & 55.6\% & 17.97 & 1.96 & 2.38 & \textbf{11.29} & 55.0\% \\
CLAP & \textbf{21.92} & 1.57 & 1.35 & 10.01 & \textbf{58.3\%} & 17.79 & 1.82 & 2.32 & 11.02 & 59.1\% \\
FlanT5-base & 23.00 & 1.55 & 1.50 & 10.11 & 53.0\% & 20.05 & 2.03 & 2.50 & 10.88 & 52.9\% \\
FlanT5-large & 22.31 & \textbf{1.41} & 1.42 & 9.37 & 54.6\% & 18.09 & \textbf{1.62} & 2.35 & 10.16 & 55.6\% \\
Clap+FlanT5-large & 22.55 & 1.50 & \textbf{1.32} & 10.34 & 57.4\% & \textbf{17.59} & 1.87 & \textbf{2.25} & 10.93 & \textbf{59.5\%} \\
\bottomrule
\end{tabular}
}
\end{table*}

\begin{table*}[ht]
\centering
\caption{Subjective evaluation for all baseline models and different encoders used in Auffusion categorized by the number of events 
in the text. OVL measures the overall quality and REL shows the relevance. ACC represents the mean accuracy of audio events matching the text in multi-event conditions, indicating the fine-grained alignment between text and audio.}
\label{tab:subjective_evaluation}
\scalebox{0.87}{
\begin{tabular}{@{}l|p{1cm}p{1cm}|cc|cc|ccc}
\toprule
\multirow{2}{*}{Model} & \multicolumn{2}{|c}{\textbf{All Event}} & \multicolumn{2}{|c}{\textbf{Single Event}} & \multicolumn{2}{|c}{\textbf{Two Events}} & \multicolumn{3}{|c}{\textbf{Multi Events}} \\
 & OVL$\uparrow$ & REL$\uparrow$ & REL$\uparrow$ & CLAP$\uparrow$ & REL$\uparrow$ & CLAP$\uparrow$ & REL$\uparrow$ & CLAP$\uparrow$ & ACC$\uparrow$ \\ \midrule
Groundtruth        & 71.56 & 74.01 & 73.70 & 50.9\% & 75.50 & 51.5\% & 72.85 & 48.7\% & 84.6\% \\ \midrule
AudioGen-v2-medium & 63.86 & 59.80 & 59.75 & 47.2\% & 58.50 & 45.9\% & 61.15 & 48.3\% & 68.5\% \\
AudioLDM-L-full    & 60.33 & 57.36 & 58.00 & 51.1\% & 60.40 & 50.8\% & 53.70 & 51.1\% & 53.2\% \\
AudioLDM2-large    & 65.53 & 59.23 & 61.50 & 50.6\% & 60.20 & 48.3\% & 56.00 & 46.9\% & 61.4\% \\
Tango-Full         & 67.78 & 65.05 & 63.75 & 48.2\% & 67.75 & 52.3\% & 62.65 & 53.6\% & 69.6\% \\ \midrule
Auffusion-w-clip   & 69.36 & 70.25 & 70.65 & 52.5\% & 73.45 & 53.2\% & 66.65 & 55.3\% & 73.9\% \\
Auffusion-w-clap   & 69.76 & 67.76 & 68.95 & \textbf{55.6\%} & 71.25 & \textbf{57.5\%} & 63.10 & \textbf{58.9\%} & 71.0\% \\
Auffusion-w-flant5 & \textbf{70.13} & 70.65 & 69.55 & 51.1\% & \textbf{74.60} & 53.0\% & 67.80 & 54.3\% & 73.3\% \\
Auffusion-w-clap-flant5 & 69.80 & \textbf{71.86} & \textbf{72.10} & 55.3\% & 73.90 & 56.8\% & \textbf{69.60} & \textbf{58.9\%} & \textbf{74.1\%} \\ 
\bottomrule
\end{tabular}
}
\end{table*}

\subsection{Analysis}
\label{sec:analysis}

\textbf{Effect of Text Encoder.} To assess the performance of different text encoders and explore the effectiveness of a dual text encoder approach in TTA applications, we compared several encoder options including CLIP, CLAP, FlanT5-base, FlanT5-large, and a combined CLAP and FlanT5-large encoder. The original SDv1.5 uses the CLIP L/14 model, trained on text-image pairs, while AudioLDM employs the CLAP model, trained on text-audio pairs. Tango suggests that FlanT5, an instruction-tuned LLM, enhances textual understanding, but lacks an encoder ablation study. Inspired by eDiff-I~\cite{ediff}, which uses an ensemble of encoders to provide multi-source information, we experimented with a combined CLAP and FlanT5-large encoder by concatenation. The results are shown in Table~\ref{tab:different_encoder}.

Our findings reveal that FlanT5-large surpasses FlanT5-base in all evaluation metrics, underscoring the importance of the text encoder's size for understanding textual captions. FlanT5-large shows results comparable to CLIP, with CLIP excelling in IS score and FlanT5-large in FAD score. This mirrors Imagen~\cite{imagen} findings, where T5-XXL matches CLIP in objective scores but exceeds smaller T5 models. Notably, the CLAP model outperforms both text encoders, especially in FAD and CLAP scores, demonstrating its advanced audio domain expertise. The elevated CLAP score may be attributed to the use of same model during training. Combining CLAP and FlanT5-large encoders leverages both acoustic and rich semantic knowledge, yielding the best overall objective performance.

\textbf{Text-Audio Alignment.} In our investigation into the impact of varying text encoders on TTA alignment—a crucial aspect of TTA tasks—we are the first to examine the cross-attention mechanisms between text and LDM outputs using method~\cite{prompt2prompt}. This approach allows us to intuitively observe the focal points of the LDM during the TTA process. However, due to the global conditioning approach employed by AudioLDM and the use of GPT for generating AudioMAE features in AudioLDM2, these models do not provide a direct correlation between text and LDM. Therefore we present a comparative visualization of cross-attention maps using various encoders within the Auffusion framework, and the Tango models who also adopt cross attention, as depicted in Figure~\ref{fig:attn_map}. For consistency, we standardize the diffusion steps to 50 and adjust all cross-attention maps to a uniform square dimension for clear comparison.

Upon comparing the attention maps of the first and second lines, it is evident that using pre-trained LDM exhibits superior distinguishability, with clear attention across almost all tokens. Notably, the ``gunshot" token within the highlighted red area is prominent, and the ``dogs" and ``gunshot" sounds in the generated audio overlap in the latter section. This suggests that the pre-trained LDM possesses advanced prior knowledge, enabling it to effectively transfer its text-image alignment capabilities to text-audio alignment tasks. Although the CLAP model achieves higher objective scores, it surprisingly produces similar attention maps for each token. Furthermore, the ``dogs" and ``gunshot" sounds are indistinguishable, occurring simultaneously in the generated audio, which indicates that the CLAP model struggles to differentiate and isolate fine-grained events or entities. We believe the reason is that CLAP model can only gather global acoustic information and have issue capturing temporal and local information in the text. Furthermore, all the objective evaluation also only focus and compute global features, therefore such inconsistency exists. Recent study~\cite{DBLP:conf/icassp/WuNBS23} corroborates that current CLAP models do not truly comprehend natural language, focusing instead on global information.

In contrast, when comparing the attention maps of the last two lines, where Tango also employs FlanT5-large as a text encoder, the Tango's attention maps appear muddled, particularly for the ``dog" token highlighted in blue area, which results in the omission of the ``dog" sound in the generated audio. Additionally, we find that Tango often produces extraneous sounds, such as unintended ``bird" noises, that are not present in the text captions. These findings highlight that Auffusion, by leveraging the robust text-image alignment capabilities of pre-trained LDMs, can generate audio that more accurately reflects the given captions.

\textbf{Performance against Number of Events.} To better assess fine-grained text-audio alignment, we evaluate performance across varying event numbers in the AudioCaps and Clotho test set. For instance, a sequence like \textit{\underline{A man talking} followed by \underline{plastic clacking} then \underline{a power tool drilling}} comprises three distinct events. We categorize the test sets into three groups: single event, two events, and multiple events (three or more), randomly selecting 20 captions from each category for generation. Human raters are asked to rate the relevance between text and audio for each group, using REL metric. Besides, we select an additional 80 samples for multiple events. We ask raters to count the number of events in the audios that accurately appear in the text, and calculate the mean accuracy denoted as ACC to reflect the fine-grained text-audio alignment. These results for baselines and Auffusion with various encoders trained on whole datasets are presented in Table~\ref{tab:subjective_evaluation}. Additionally, we used the objective CLAP score for comparison.

We assume that human raters can directly and faithfully represent true performance. We find that the CLAP score can not accurately reflect detailed alignment ability, particularly in multi-event evaluation, compared with human evaluation. Our analysis concludes that CLAP primarily extracts global features, lacking in fine-grained evaluation capacity. Additionally, we observe that AudioLDM, AudioLDM2, and AudioGen exhibit inferior performance in REL and we can tell from ACC score that they fail to generate matching audio in multi-event scenarios. This is attributed to AudioLDM using globally pooled CLAP embeddings for conditioning, while AudioLDM2 first employs an auto-regressive model (AR) to generate AudioMAE features from text, then uses these to condition the LDM. Consequently, fine-grained information is lost in AudioLDM, and the AR model in AudioLDM2 introduces error accumulation. In contrast, Tango and our Auffusion, which adopt cross-attention between text embedding sequences and LDM, demonstrate better alignment. Moreover, our findings across various encoders align with the attention map results illustrated in previous part. Despite Auffusion's combination with the CLAP encoder yielding a higher CLAP score, evaluations using ACC and REL especially in multi-event scenarios reveal that the CLAP encoder captures less fine-grained information compared to CLIP and FlanT5 encoders.

\textbf{Applications.} Leveraging our system's exceptional text comprehension capabilities and robust text-audio alignment, we demonstrate its versatile applications inspired by T2I tasks~\cite{SDEdit,palette,controlnet,lama}. These include audio style transfer, audio inpainting, and attention-based techniques such as word swap and text attention re-weighting. We demonstrate these capabilities in Appendix~\ref{sec:suppl_demo}. Our method offers a significantly more controllable and fine-grained manipulation compared to previous methods\cite{AudioLDM,AudioLDM-2,audiogen,tango}.

\section{Conclusion}
\label{sec:conclusion}

In this study, we introduce Auffusion, a text-to-audio (TTA) generation model that harnesses the robust generative capabilities and precise cross-modal alignment abilities of pretrained text-to-image (T2I) models. Our extensive objective and subjective evaluations demonstrate that Auffusion surpasses other state-of-the-art models, achieving superior performance with limited data and computational resources. Recognizing the significant impact of different encoders on cross-modal alignment in T2I, we pioneer in the TTA field by conducting comprehensive investigations and innovatively adopting cross-attention map visualization. This approach offers an intuitive evaluation of text-audio alignment. Our findings demonstrate that Auffusion exhibits an exceptional ability to generate audio that accurately aligns with text descriptions, surpassing existing methods, which further evidenced in several audio manipulations, including audio style transfer, inpainting, word swapping, and re-weighting. In the future, we aim to delve into a broader spectrum of innovative audio applications, based on the robust text-audio alignment capabilities of our system.

\newpage
{
    \small
    \bibliographystyle{ieeenat_fullname}
    \bibliography{main}
}

\clearpage
\setcounter{page}{1}

\renewcommand{\thesection}{\Alph{section}}
\setcounter{section}{0}

{   
\newpage
\onecolumn
\centering
\Large
\textbf{\thetitle}\\
\vspace{0.5em}Supplementary Material \\
\vspace{1em}
}


\section{Dataset Details}

\begin{table*}[h]
\centering
\caption{Statistics for the all datasets used in the this paper.}
\begin{tabular}{lllll}
\toprule
Dataset            & Samples & Hours (Original) & Hours (Processed) & Source \\ \hline
AudioCaps          & 41K     & 110h       & 110h &  \cite{audiocaps}      \\
MACS               & 3930    & 11h        & 11h  &  \cite{macs}      \\
Clotho             & 5929    & 37h        & 37h  &  \cite{clotho}      \\
ESC50              & 2000    & 3h         & 3h   &  \cite{esc50}      \\
UrbanSound8K       & 8266    & 9h         & 9h   &  \cite{urbansound}      \\
Music Instruments  & 4587    & 11h        & 10h  &  \cite{tango}    \\
GTZAN              & 2997    & 8h         & 8h   &  \cite{gtzan}      \\
BBC Sound Effects  & 31K     & 997h       & 232h &  \href{https://sound-effects.bbcrewind.co.uk/}{https://sound-effects.bbcrewind.co.uk/}    \\
FreeSound          & 206K    & 6246h      & 1283h &  \href{https://freesound.org/}{https://freesound.org/}        \\
AudioSet\_SL       & 108K    & 296h       & 296h &  \cite{audioset-sl}      \\
SoundBible         & 612     & 4h         & 3h   &   \href{https://soundbible.com/}{https://soundbible.com/}     \\
\midrule
Total              & 402K    & 7732h      & 1990h &  -      \\
\bottomrule
\end{tabular}
\label{tab:suppl_dataset}
\end{table*}

As indicated in Table~\ref{tab:suppl_dataset}, we collect a large-scale audio-text dataset comprising approximately 0.47M audio samples, amounting to a total duration of about 7.7K hours. This dataset encompasses a diverse range of sounds, including musical instruments, sound effects, human voices, and sounds from nature and everyday life. We only utilize the first 30 seconds of each audio sample for long duration audio and exclude any samples shorter than 1 second. Consequently, our model is trained on approximately 0.4M audio samples, collectively amounting to about 2K hours.

\section{Experiment Details}

\subsection{Vocoder} 

In this work, we adopt HiFi-GAN vocoder~\cite{hifigan} as a converter from VAE decoder output to finally generated audio. It is widely used for speech waveform and audio sample reconstructed from mel-spectrogram. However, the default and pretrained HiFi-GAN\footnote{https://github.com/jik876/hifi-gan} use 80 mel filter bands, 256 hop size and trained on 22050 sample rate speech, resulting in (1,80,882) mel-spectogram for 10-second audio, which does not suit VAE input requirements. Therefore we train our own HiFi-GAN vocoder using 256 mel filter bands, 1024 window length, 2048 FFT, and 160 hop size, resulting in (1,256,1024) mel-spectrograms. Then we repeat by channel and convert grayscale image to RGB image to match VAE encoder input. The (3,256,1024) image is then 8x downsampled by VAE as (4,32,128) latent features sent to LDM. We train this vocoder using AdaW optimizer with 2e-4 learning rate and 16 batch size on one A6000 GPU. We release this pretrained vocoder in our open-source implementation.

\subsection{Configuration} 

We utilize the pretrained Stable Diffusion v1.5\footnote{https://huggingface.co/runwayml/stable-diffusion-v1-5}, including its VAE and 860M-parameter U-Net. we freeze the text encoder and finetune U-Net using AdamW optimizer with a 3e-5 learning rate and a constant scheduler, at a 20 batch size for 100K steps in both Auffusion and Auffusion-Full setups. For GPU memory efficiency, we use xformer~\cite{xformers} and mixed precision, and our model can be trained only taking a total of 48 hours on one A6000 GPU. In comparison, AudioGen~\cite{audiogen} utilizes 64 A100 GPU with a batch size of 256. AudioLDM~\cite{AudioLDM} use one A100 GPU for 1.5M train steps and AudioLDM2~\cite{AudioLDM-2} use 8 A100 GPU for the same steps. Tango use 4 A6000 GPU for 200K steps.

\subsection{Riffusion Re-implementation} 

Riffusion~\cite{riffusion} is originally trained on music datasets for only 5-second sound clips using a 44100 sample rate, a 441 hop size, and 512 mel filter bands. We changed the sample rate to 16000 and the hop size to 160 to match audio generation setup. After converting audio to a mel-spectrogram, Riffusion simply applies min-max normalization to each individual mel-spectrogram and quantizes it into an image. Therefore, it is a non-reversible process that leads to information loss. To convert back to audio, Riffusion uses the Griffin-Lim~\cite{griffin-lim} algorithm, which is not sensitive to initial data range and iteratively estimates the missing phase information. However, the quality is not comparable to that of deep-learning-based vocoders. We use the same training setup to re-implement Riffusion with same whole datasets.

\section{Effect of Guidance Scale and Inference Steps}

The number of inference steps and the classifier-free guidance scale are of crucial importance for sampling from latent diffusion models. We report the effect of varying these parameters on audio generation using AudioCaps test set in Table~\ref{tab:suppl_guidance_steps}. On the left, with a fixed guidance scale of 7.5, we explore inference steps ranging from 10 to 200. Unlike previous studies~\cite{tango,AudioLDM} using 200 steps, we find that our Auffusion model have competent performance even with fewer inference steps, suggesting strong generative capabilities. Therefore we choose using 100 inference steps for Auffusion model. On the right, we fix the steps at 100 and adjust the guidance scale. We find that guidance 5 has the best FD and FAD score, and guidance 10 excels in IS and CLAP score, therefore we choose balanced guidance scale 7.5 for Auffusion model to generate audio.

\begin{table*}[h]
\centering
\caption{Effect on the objective evaluation metrics with a varying number of inference steps and classifier-free guidance scale.}
\label{tab:suppl_guidance_steps}
\scalebox{0.88}{
\begin{tabular}{ccccccc|ccccccc}
\toprule
\multicolumn{7}{c|}{\textbf{Varying Steps}} & \multicolumn{7}{|c}{\textbf{Varying Guidance}} \\
\midrule
Guidance & Steps & FD$\downarrow$ & FAD$\downarrow$ & KL$\downarrow$ & IS$\uparrow$ & CLAP$\uparrow$ & Steps & Guidance & FD$\downarrow$ & FAD$\downarrow$ & KL$\downarrow$ & IS$\uparrow$ & CLAP$\uparrow$ \\ 
\midrule
\multirow{5}{*}{7.5} &10 & \textbf{21.15} & 2.48 & 1.54 & 7.89 & 48.9\% & \multirow{5}{*}{100} & 1 & 32.32 & 5.17 & 2.47 & 4.56 & 32.2\% \\
& 25  & 22.52 & 2.03 & 1.36 & 9.88  & 55.3\% & & 2 & 23.44 & 2.86 & 1.63 & 6.4 & 47.1\% \\
& 50  & 23.35 & 2.01 & \textbf{1.35} & 10.09 & 55.5\% &  & 5 & \textbf{21.24} & \textbf{1.72} & 1.37 & 9.29 & 54.9\% \\
& 100 & 23.44 & 1.96 & 1.36 & \textbf{10.21} & \textbf{55.6\%} &  & 7.5 & 23.44 & 1.96 & \textbf{1.36} & 10.21 & 55.6\% \\
& 200 & 23.34 & \textbf{1.92} & 1.36 & 10.20 & \textbf{55.6\%} &  & 10  & 24.69 & 2.31 & \textbf{1.36} & \textbf{10.50} & \textbf{55.9\%} \\
\bottomrule
\end{tabular}
}
\end{table*}

\section{Evaluation}

\begin{figure*}[h]
  \centering
   \includegraphics[width=0.7\linewidth]{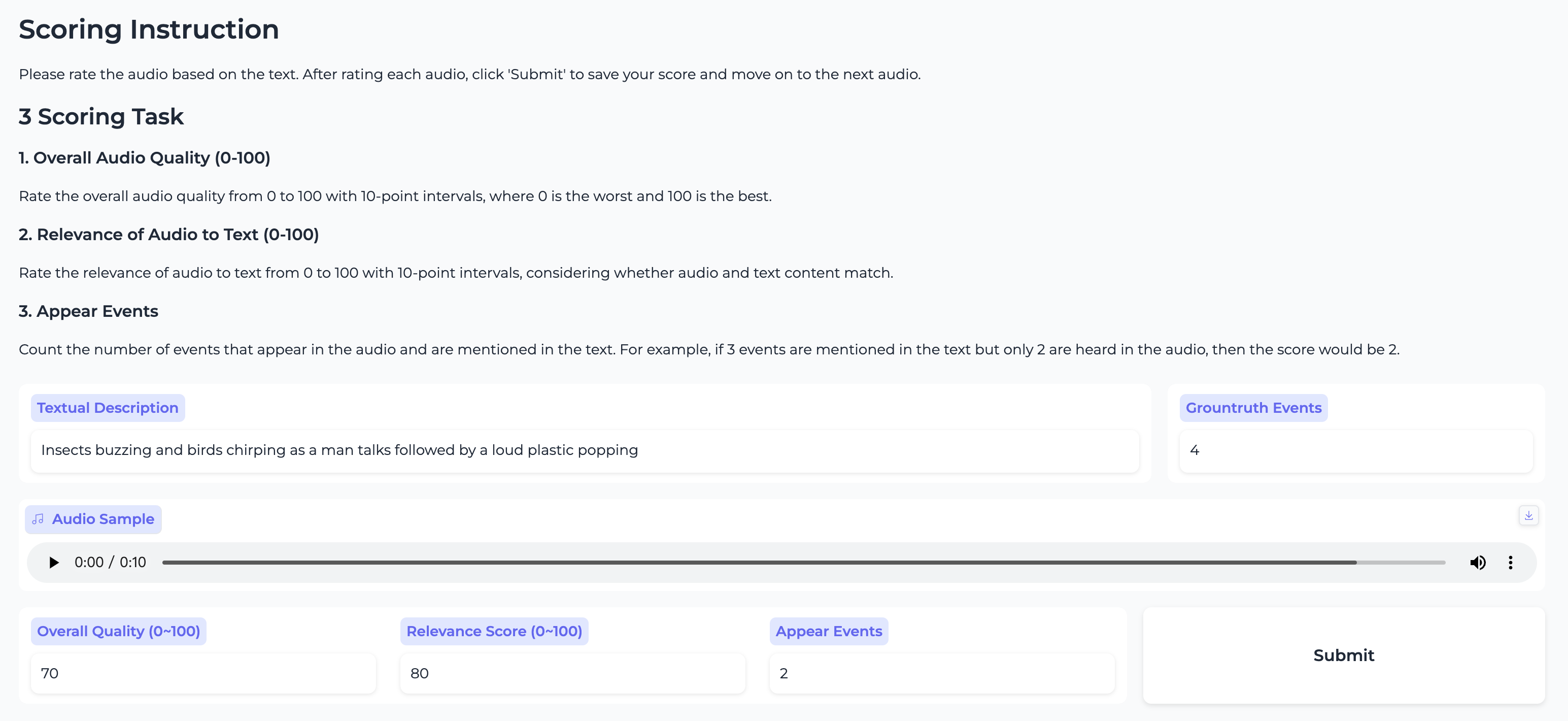}
   \caption{Screenshot of subjective evaluation.}
   \label{fig:suppl_scoring_demo}
\end{figure*}

Despite many objective evaluation method exits, they can only assess global performance. Subjective method is a much more direct approach and can be tailored to specific needs. We design three scoring tasks to evaluate performance. We first ask human raters to evaluate the overall qaulity (OVL) from 0 to 100 with 10-point intervals. Then they need to rate the relevance (REL) to the text caption, evaluate the global text-audio alignment. Finally, we ask testers to count the number of events that appear in the audio to assess the fine-grained text-audio alignment. Our designed subjective evaluation is shown in Figure~\ref{fig:suppl_scoring_demo}.

\newpage

\newpage

\section{Demos}
\label{sec:suppl_demo}

\subsection{Text-to-Audio Generation}
\label{subsec:suppl_TTA}

\begin{figure*}[h]
  \centering
   \includegraphics[width=\linewidth]{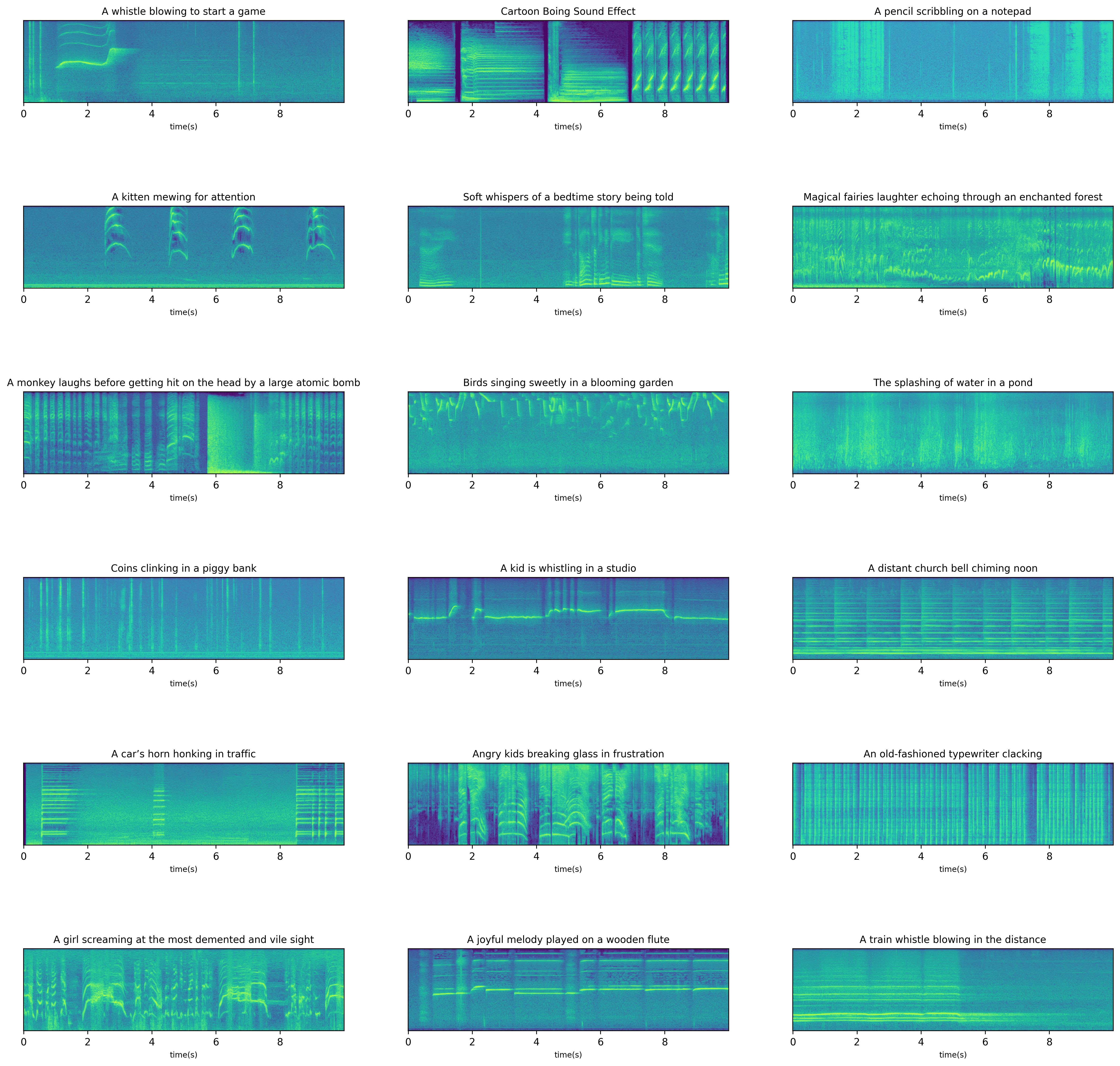}
   \caption{Demo of audio generation with the Auffusion-Full model.}
   \label{fig:suppl_demo}
\end{figure*}

\newpage

\subsection{Audio Style Transfer Examples}
\label{subsec:suppl_style_transfer}

\begin{figure*}[h]
    \centering
    \vspace{0.5cm}  
    \includegraphics[width=0.9\linewidth]{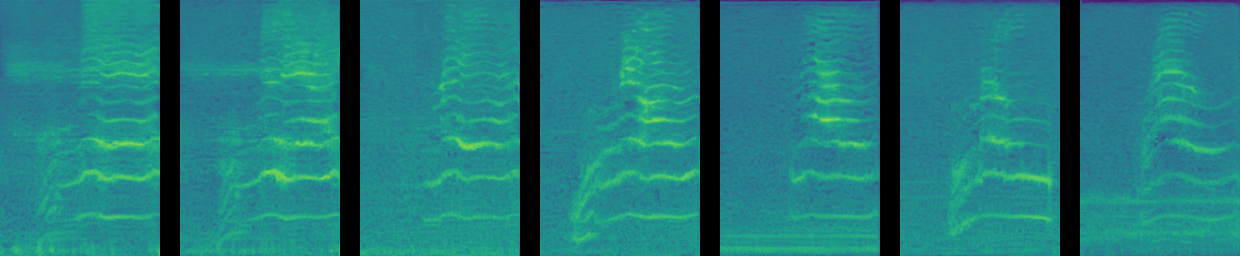}
    \caption{Audio style transfer gradually from \textbf{baby crying} to \textbf{cat meowing}.}
    \label{fig:suppl_style_1}
\end{figure*}

We show the audio style transfer ability of our Auffusion. We adopt the similar image-to-image manipulation method first introduced in T2I task by using shallow reverse process~\cite{SDEdit} to audio domain. In the figures below, we show the original audio samples on the left, and six transferred audios guided by textual descriptions using different starting points of the reverse process $n$. As $n$ increases, more noise will be added to the original audio. Diffusion model will pay more attention on text guidance and the generated audio will become less similar to the original one. When $n=1$, the added noise is at its maximum, and information from the original audio will not be retained. We gradually increase $n$ and set $n=0.7$ for the last generated sample. For instance, the original sound of a \textbf{baby crying} gradually transitions into the sound of a \textbf{cat meowing} in Figure~\ref{fig:suppl_style_1}. 

\begin{figure*}[h]
    \centering
    \vspace{0.5cm}  
    \includegraphics[width=0.9\linewidth]{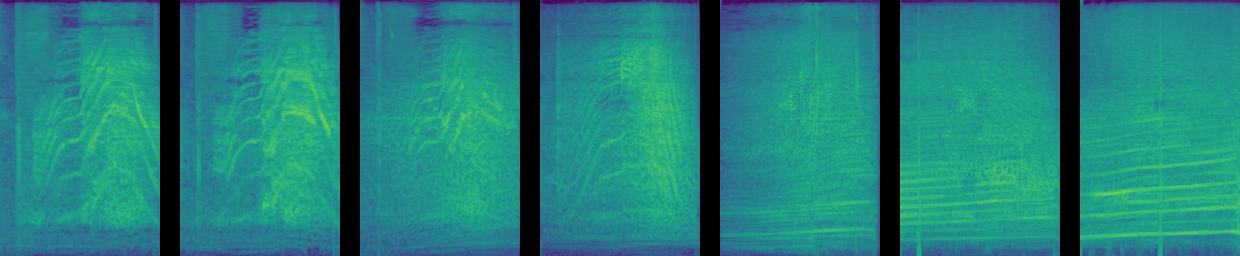}
    \caption{Audio style transfer gradually from \textbf{cat screaming} to \textbf{car racing}.}
    \label{fig:suppl_style_2}
\end{figure*}

\begin{figure*}[h]
    \centering
    \vspace{0.5cm}  
    \includegraphics[width=0.9\linewidth]{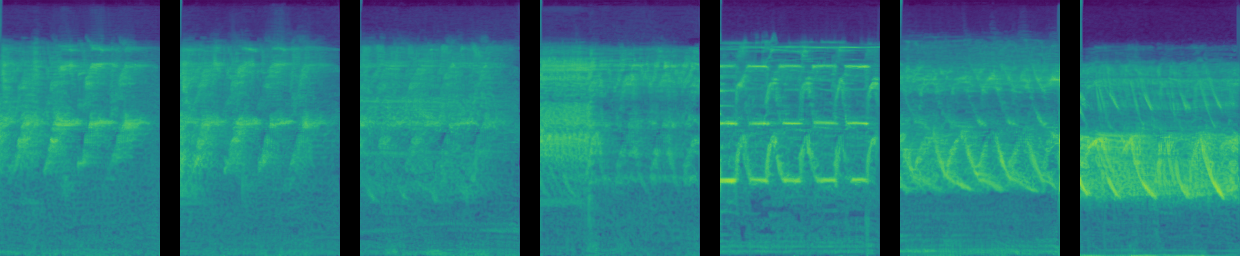}
    \caption{Audio style transfer gradually from \textbf{bird chirping} to \textbf{ambulance siren}.}
    \label{fig:suppl_style_3}
\end{figure*}

\newpage

\subsection{Audio Inpainting Examples}
\label{subsec:suppl_audio_inpaint}

In the figure below, we demonstrate the audio inpainting capability of our Auffusion model. Each audio sample is extended to a duration of 10 seconds. In the first row of each example, we present the groundtruth samples. For the unprocessed samples, we mask a segment from 2.5 to 7.5 seconds in the original audio and use these masked samples as input for inpainting. The inpainting results are produced using the same text prompt as the original groundtruth sample. We observe that our model can comprehend the textual description and audio context, thereby generating appropriate content for the masked segment.

\begin{figure*}[h]
  \centering
   \includegraphics[width=1\linewidth]{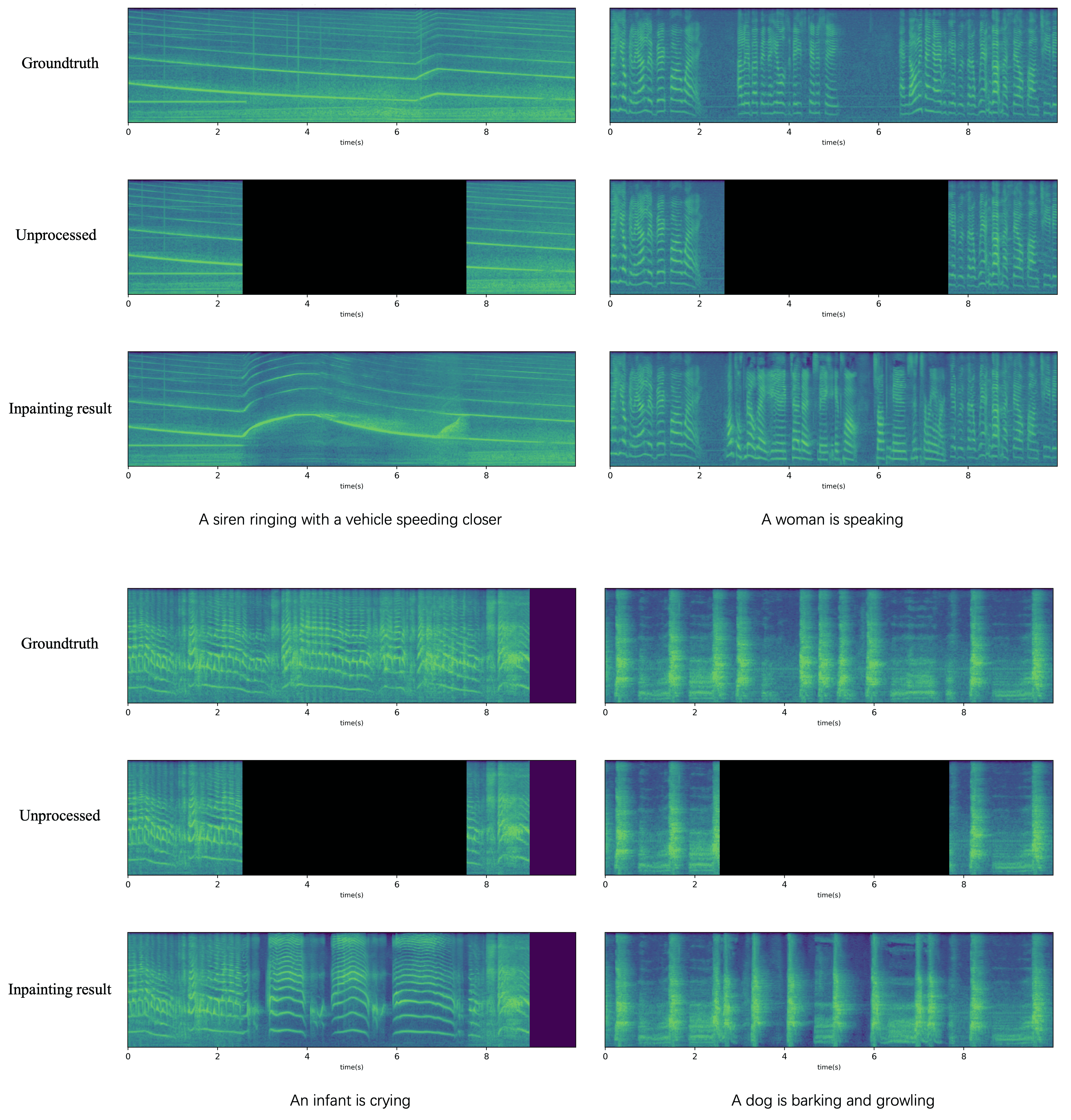}
   \caption{The examples of audio inpainting ability of Auffusion.}
   \label{fig:suppl_inpainting}
\end{figure*}

\newpage

\subsection{Word Swap Examples}
\label{subsec:suppl_replacement}

Showing the replacement ability of attention map in TTA task. In below cases, we swap tokens in the original prompt with others. By changing \textbf{huge} to \textbf{small}, we observe that the sound effect in the corresponding part changes to a less echoic and clearer sound. By replacing \textbf{gunshots} to \textbf{speech}, the corresponding sound is replaced with a human voice.

\begin{figure*}[h]
  \centering
   \includegraphics[width=0.9\linewidth]{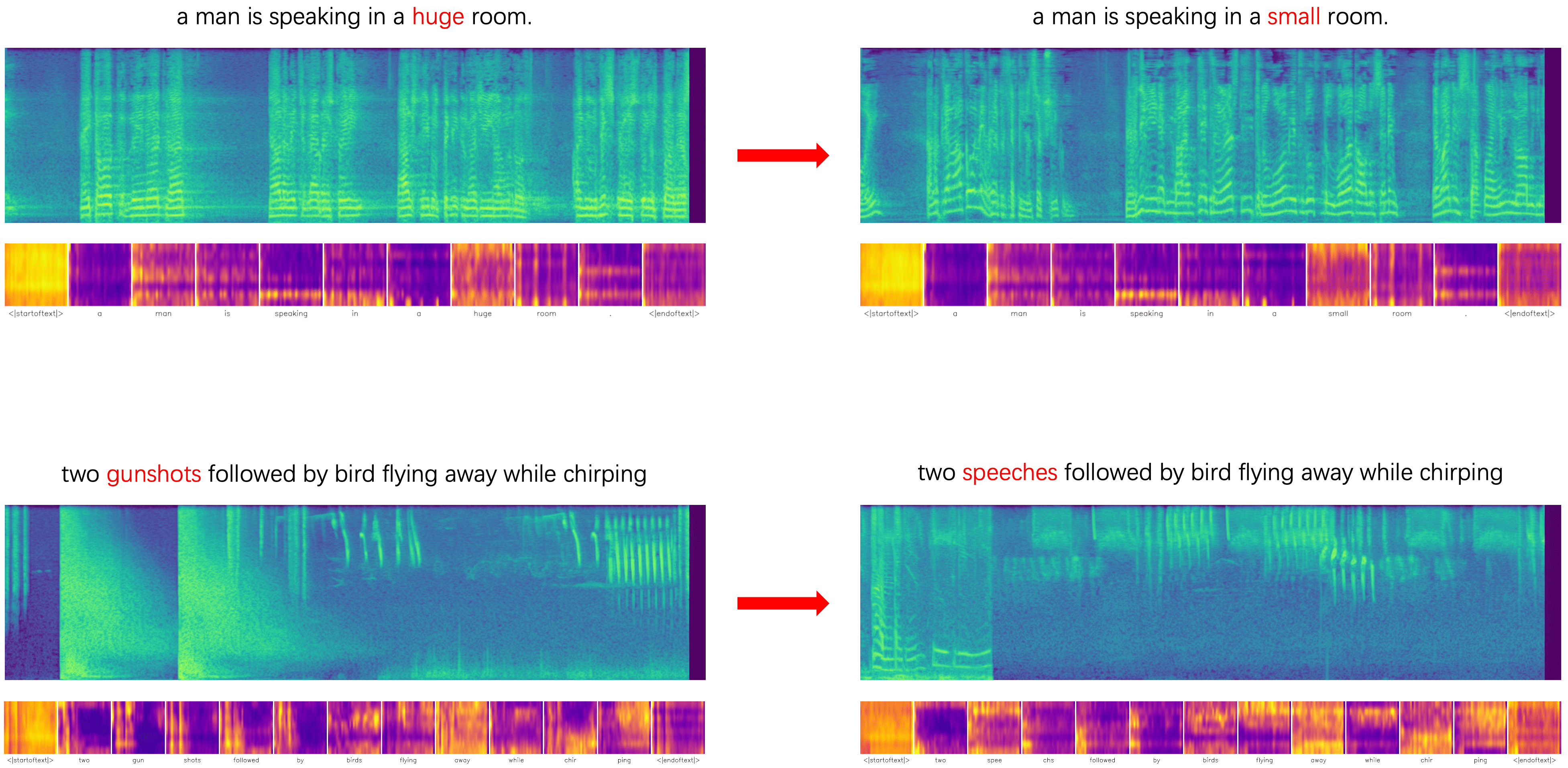}

   \caption{Demo of word swapping manipulation.}
   \label{fig:suppl_word_swap}
\end{figure*}

\vspace{-0.2cm}
\subsection{Attention Re-weighting Examples}
\label{subsec:suppl_reweight}

Showing the re-weighting ability of attention map in TTA task. By increasing the cross attention of specific words (marked with an arrow), we control the effect only on specific words without significant change the image. We find that increasing the weight on the verb \textbf{chopping} enhances the frequency of the action sound, while amplifying the adjective \textbf{huge} affects the sound effect.

\begin{figure*}[h]
  \centering
   \includegraphics[width=0.9\linewidth]{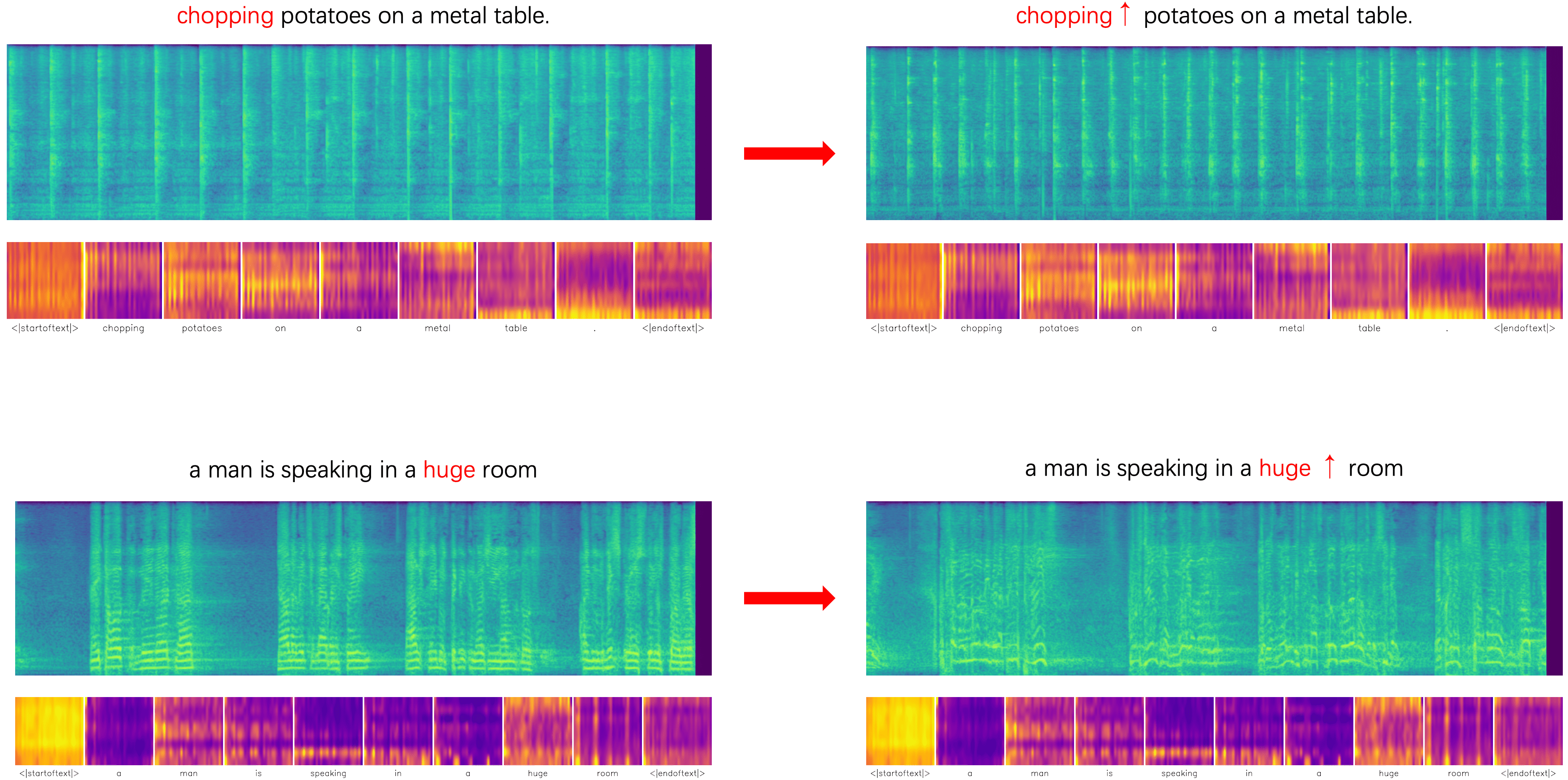}
   \caption{Demo of attention re-weighting manipulation.}
   \label{fig:suppl_reweight}
\end{figure*}


\end{document}